\documentclass{article}
\usepackage{booktabs}
\usepackage{spconf,amsmath,graphicx}
\usepackage{xcolor}
\usepackage{amsmath}
\usepackage{amssymb}
\usepackage[caption = false]{subfig}
\usepackage{cite}
\usepackage{comment}
\usepackage{hyperref}

\newcommand{\Am}[1]{\mathbf{A}_{#1}}
\newcommand{\Bm}[1]{\mathbf{B}_{#1}}
\newcommand{\Cm}[1]{\mathbf{C}_{#1}}

\title{DASS: Distilled Audio State Space Models \\Are Stronger and More Duration-Scalable Learners}
\name{Saurabhchand Bhati$^{1}$, Yuan Gong$^{1}$, Leonid Karlinsky$^{2,3}$, Hilde Kuehne$^{3,4}$, Rogerio Feris$^{2,3}$, James Glass$^{1}$}
\address{$^{1}$MIT, USA, $^{2}$IBM Research AI, USA, $^{3}$MIT-IBM Watson AI Lab, USA, $^{4}$University of Bonn, Germany \\
\small{sbhati@mit.edu}}
\begin{document}

\maketitle

\begin{abstract}

State-space models (SSMs) have emerged as an alternative to Transformers for audio modeling due to their high computational efficiency with long inputs. While recent efforts on Audio SSMs have reported encouraging results, two main limitations remain: First, in 10-second short audio tagging tasks, Audio SSMs still underperform compared to Transformer-based models such as Audio Spectrogram Transformer (AST). Second, although Audio SSMs theoretically \emph{support} long audio inputs, their \emph{actual performance} with long audio has not been thoroughly evaluated. To address these limitations, in this paper, 1) We applied knowledge distillation in audio space model training, resulting in a model called Knowledge \underline{D}istilled \underline{A}udio SSM (DASS). To the best of our knowledge, it is the first SSM that outperforms the Transformers on AudioSet and achieves an mAP of 48.9; and 2) We designed a new test called Audio \underline{N}eedle \underline{I}n \underline{A} \underline{H}aystack (Audio NIAH). We find that DASS, trained with only 10-second audio clips, can retrieve sound events in audio recordings up to 2.5 hours long, while the AST model fails when the input is just 50 seconds, demonstrating SSMs are indeed more duration scalable. Code: \href{https://github.com/Saurabhbhati/DASS}{Github}, \href{https://huggingface.co/saurabhati/DASS_small_AudioSet_48.9}{HuggingFace}

\end{abstract}

\section{Introduction}

Transformers have become the primary choice for modeling speech, text, and image data~\cite{vaswani2017attention,radford2023robust,dosovitskiy2020image,touvron2021training,touvron2023llama}. Over the last few years, audio classification systems have gradually shifted from CNN-based models~\cite{kong2020panns,gong2021psla,miyazaki2020convolution,kong2020sound} to Transformer-based models~\cite{gong2021ast,chen2022hts,koutini2021efficient,huang2022masked}. Transformer models, based on the attention mechanism, enjoy a larger receptive field; however, they have quadratic computational complexity, which makes them inappropriate for processing long sequences~\cite{gong2021ast,chen2022hts}. Recently, state-space models (SSM)\cite{gu2021efficiently,gu2023mamba} have emerged as an alternative to Transformer-based models for sequence modeling, providing a more efficient approach, especially for long sequences. Concurrent to our work, there have been approaches that adapt SSM-based models for audio classification tasks such as AuM\cite{hamza2024audio}, Audio Mamba~\cite{lin2024audio}, and SSAMBA~\cite{shams2024ssamba}.

\begin{figure}[th!]
    \subfloat[Proposed knowledge distillation framework]{
    \hspace{-1pt}
    \includegraphics[width=0.6\columnwidth]{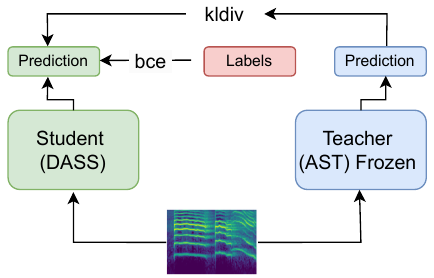}
    }
    \subfloat[Performance on AudioSet]{
    \includegraphics[width=0.37\columnwidth]{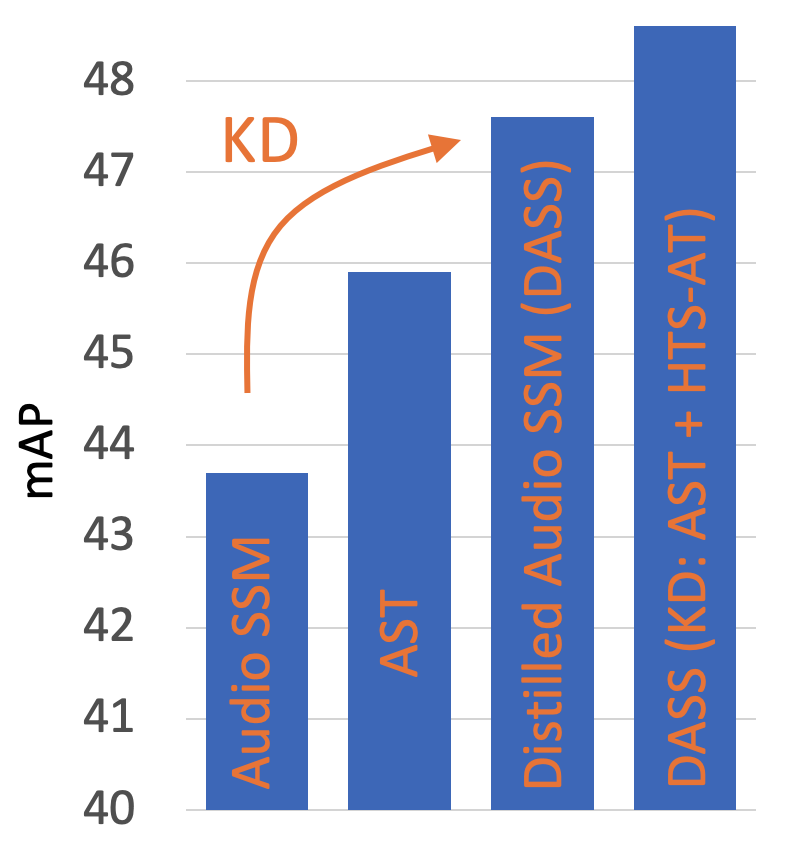}
    }  \\
    \centering
    \subfloat[Audio NIAH performance vs. input audio length]{
    \includegraphics[width=0.7\columnwidth]
    {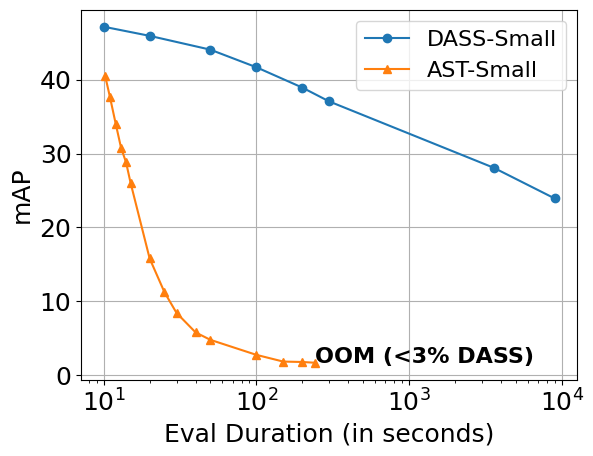}}
    \caption{(a) We apply knowledge distillation in audio SSM training using AST as the teacher. (b) The trained distilled audio SSM (DASS) outperforms the AST teacher. Using an ensemble of teachers(AST~\cite{gong2021ast} + HTS-AT~\cite{chen2022hts}) further boosts the performance. (c) DASS can take up to 2.5 hours of audio input ($>$ 30$\times$ longer than AST) on a single A6000 GPU while maintaining performance well. Note that the horizontal axis is in log scale.}
    \label{fig:DASS_overview}
    \vspace{-10pt}
\end{figure}

In this work, we have made two major technical contributions: First, the performance of existing audio SSM models~\cite{hamza2024audio,lin2024audio,shams2024ssamba} still falls short of transformer-based models on audio classification tasks. We found that knowledge distillation is an effective technique to enhance state-space models, enabling them to \emph{outperform} their Transformer counterparts. Specifically, when trained with knowledge distillation using AST as the teacher, our DASS model achieved an mAP of 47.6, outperforming the teacher AST model (45.9 mAP) with a 1.8 times smaller model size. Using an ensemble of teacher i.e. AST~\cite{gong2021ast} and HTS-AT~\cite{chen2022hts}, DASS model achieves significantly higher mAP of 48.9 while maintaining the smaller model size. 

Second, while recent SSM-based audio model works~\cite{hamza2024audio,shams2024ssamba} have shown theoretical GPU memory and inference speed-up benefits over transformer-based models, they have not measured the \emph{actual} performance of these models at longer input lengths. To fill this gap, we designed a new benchmark called Audio Needle In A Haystack (Audio NIAH) to measure the long audio inference performance of audio classification models trained with only short audio segments. Specifically, we insert the needle, a 10-second audio event, into a larger haystack and evaluate how well the model can classify the needle audio event. In our experiment, we surprisingly found DASS to be significantly stronger than AST in long audio inference. Specifically, when both models are trained with only 10-second audios, the performance of AST models drops to less than 5 mAP when the input is 50 seconds, which is $<$ 12\% of the performance for 10-second input, while DASS's performance is 45.5 mAP (96\%) in the same setting. On a single A6000 GPU, DASS can take up to 2.5-hour long audio input and still maintain 62\% of its performance compared to a 10-second input.

\section{Related Works}
Initial approaches for audio event classification relied on the CNN~\cite{kong2020panns,gong2021psla} or CNN-transformer hybrid models~\cite{miyazaki2020convolution,kong2020sound}. The Audio Spectrogram Transfomer (AST)~\cite{gong2021ast} proposed a convolutional free, purely attention-based model that outperformed existing models on the audio-event classification task. Hierarchical Token-Semantic Audio Transformer (HTS-AT)~\cite{chen2022hts} proposed swin transformer-based models which reduced the model size. 
Patchout faSt Spectrogram Transformer (PaSST)~\cite{koutini2021efficient} proposed a patchout method to reduce the input sequence length to the transformer model. Both these models outperform AST both in terms of performance and compute requirements.  Audio-masked autoencoder (Audio-MAE)~\cite{huang2022masked} proposed an MAE based self-supervised learning framework and achieved state-of-the-art performance for audio classification across a wide variety of datasets.  

For transformers, the compute and memory complexity increase quadratically with the input length. SSMs offer an alternate solution to transformers with linear complexity with input lengths.
Gu et al.~\cite{gu2021efficiently} showed the potential of SSMs for modeling long input sequences. Gu et al.~\cite{gu2023mamba} proposed a data-dependent selective-scan state-space block and extended the SSM framework. Their model outperformed Transformer models on large-scale natural language data, thus sparking the rise of SSM models as a generic sequence modeling backbone~\cite{mehta2022long,zhu2024vision,liu2024vmamba}. 
Zhu et al.~\cite{zhu2024vision} proposed a bidirectional SSM, which combines forward and backward SSMs to improve the modeling capabilities for vision data. Liu et al.~\cite{liu2024vmamba} further extended the state-space models and proposed a 2D-Selective-Scan algorithm (SS2D) for visual inputs. SS2D scans the input in four directions: left-to-right, right-to-left, top-to-bottom, and bottom-to-top. Each sequence is processed by a selective-scan state-space block and outputs are merged to create the final output.  

Concurrently, some approaches have adapted the SSM to audio-event classification: AuM~\cite{hamza2024audio}, Audio Mamba~\cite{lin2024audio} and SSAMBA~\cite{shams2024ssamba}. AuM and Audio Mamba adapt Vision Mamba and VMamba for audio and show remarkable performance and computational efficiency. SSAMBA explores self-supervised SSM learning for audio. The closest work to ours is Audio Mamba~\cite{lin2024audio} and the major difference is that 1) we use knowledge distillation and achieve state-of-the-art performance on audio event classification, and 2) we propose the NIAH task and evaluate SSM duration scalability.

\begin{figure*}[ht!]
    \subfloat[Architecture of the state-space model in DASS]{
    \hspace{-1pt}
    \includegraphics[ height=2.0in,width=1.55\columnwidth]{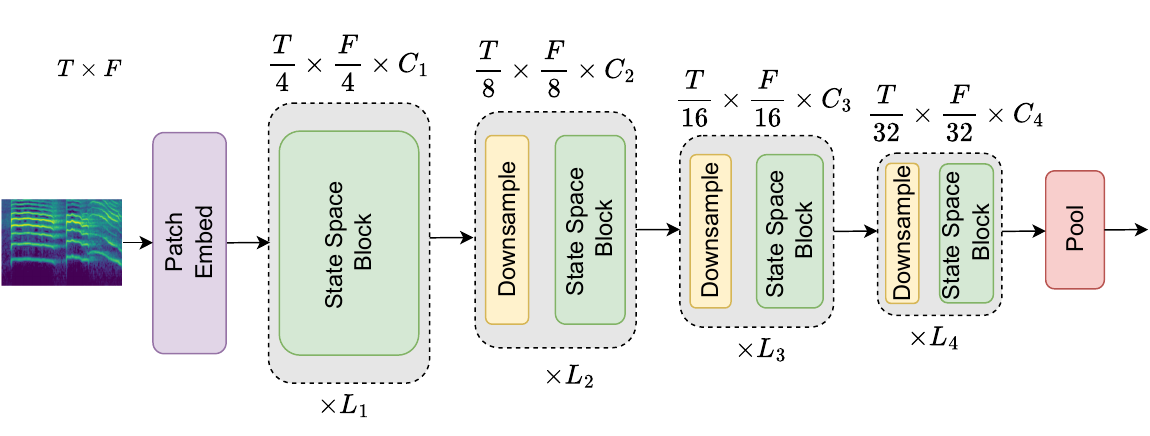}
    }  
    \subfloat[State space block]{
    \includegraphics[ height=2in,width=0.38\columnwidth]{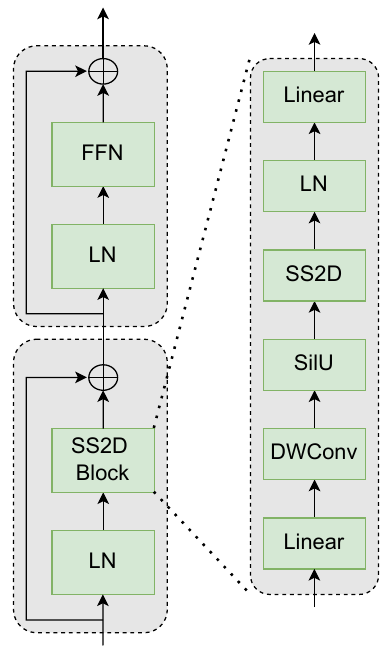}
    }
    \caption{State-space architecture in the DASS model. The state space block is analogous to the transformer block. The down sample module reduce the spatial dimension and increases the number of channels. 
    FFN: Feed Forward Network; LN: Layer Norm; SS2D: Spatial Squeeze and Dimensional, DWConv: DepthWise Convolution}
    \label{fig:DASS_arch}
    \vspace{-5pt}
\end{figure*}

\section{Distilled Audio State-Space Models}
\subsection{State-Space Models}
Structured state space sequence models (S4)~\cite{gu2021efficiently} are inspired by classical state-space models such as Kalman filters and are broadly related to recurrent neural networks (RNNs) and convolutional neural networks (CNNs). The continuous state-space models map a 1-D function or sequence $ x(t) \in \mathbb{R} \rightarrow y(t) \in \mathbb{R}$ through a hidden state $h(t) \in \mathbb{R}^{N}$ via linear ordinary differential equations as follows:
\begin{align}
    & \mathbf{h'(t)} = \Am{}\mathbf{h(t)} + \Bm{}x(t), \\
    & y(t) = \Cm{}\mathbf{h(t)}   
\end{align}
where $ \mathbf{A} \in \mathbb{R}^{N\times N}$ is called the evolution parameter and $ \mathbf{B} \in \mathbb{R}^{N\times 1}, \mathbf{C} \in \mathbb{R}^{1\times N}$ are called the projection parameters.

To adapt the state-space model to neural models, a discretization method is applied. The commonly used method for discretization is zero-order hold (ZOH) which uses a timescale parameter $\Delta$ to transform the continuous parameters $\Am{}$, $\Bm{}$ to discrete parameters $\overline{\Am{}}, \overline{\Bm{}}$ as follows:
\begin{align}
    &\bar {\Am{}} = \exp(\Delta \Am{}), \\
    &\bar {\Bm{}} = (\Delta \Bm{})^{-1}(\exp(\Delta \Am{}) - \mathbf{I})\Delta \Bm{}
\end{align}

After discretization, the state-space equations can be rewritten as:
\begin{align}
    & h_{t} = \overline{\Am{}}h_{t-1} + \overline{\Bm{}} x_{t} \\
    & y_{t} = \Cm{} h_{t}
\end{align}

The output at the current hidden state only depends on the previous hidden state. This view of state-space models can be considered analogous to RNNs. Since SSM parameters are not time-dependent, an SSM can also be viewed as a convolution $ y_{t} = (x_{0},x_{1},...,x_{t}) * (\Cm{}\overline{\Bm{}},\Cm{}\overline{\Am{}\Bm{}},...,\Cm{}\overline{\Am{}}^{M-1}\overline{\Bm{}}) = \mathbf{x} * \overline{\mathbf{K}} $ where $\overline{\mathbf{K}}$ is the called the global convolutional kernel. 

One major advantage of SSMs is that we can view them as either CNNs or RNNs depending on the task. During inference, the S4 can be viewed as an RNN, allowing faster inference and unbounded context.  During training, a convolutional view is used to enable parallel training like CNNs. 

However, the linear time-invariant nature of these models does not capture contextual information well and performs poorly on content-based reasoning tasks. 
To tackle the limitations of SSMs, Gu et al.~\cite{gu2023mamba} proposed a parametrization method to make the timescale parameter input-dependent and proposed selective scan S4. However, the state-parameters are now input dependent and the convolutional view can not be used. This poses a challenge for efficient computation. The recurrent view can still be derived and a hardware-aware parallel algorithm is used to efficiently compute the output.

\subsection{DASS}

The overview of our DASS model is shown in Figure~\ref{fig:DASS_overview} and Figure~\ref{fig:DASS_arch}  shows the detailed architecture of the state-space student model. The model can be subdivided into four groups and each group consists of a state-space block and all the groups except the first one also have a patch merging based downsampling layer. The model progressively reduces the spatial dimensions and increases the number of features. A pooling method generates the final embedding for the spectrogram which is then passed into a classifier to generate the final output of the model. 

Specifically, the model takes a two-dimensional spectrogram $ X \in \mathbb{R}^{T \times F} $ as input then a patch embedding layer extracts two-dimensional feature patches with spatial dimensions $\frac{T}{4} \times \frac{F}{4} \times C_1$. The first group processes the features at this scale and generates the output at the same scale. The next group downsamples the features to $\frac{T}{8} \times \frac{F}{8} \times C_2$ dimensions. This continues for two blocks and ultimately we obtain features with spatial dimension $\frac{T}{32} \times \frac{F}{32} \times C_4$. We use a pooling method to summarize information into a single embedding which is then passed into a linear layer to generate the final output.  

We want to leverage the existing transformer-based models to train and boost the performance of SSM. We use knowledge distillation from a transformer-based teacher model (AST) to distill knowledge into an SSM-based student (DASS). We pass the same input spectrogram to the student (DASS) and the teacher (AST) and generate the output from the two models. The student model is trained to mimic the output of the teacher model and predict the ground truth labels. The overall loss for the model is $
    \mathcal{L} = 0.5(\mathcal{L}_{bce}(y,\hat{y}_{stu}) + \mathcal{L}_{kldiv}(\hat{y}_{teach},\hat{y}_{stu}))$,
where $\mathcal{L}_{bce}$ is the binary cross-entropy loss between the output of the student DASS model $\hat{y}_{stu}$ and the ground truth labels $y$ and $\mathcal{L}_{kldiv}$ is the KL-divergence between $\hat{y}_{stu}$ and the output of teacher model $\hat{y}_{teach}$. 

\section{Experiments}

We evaluate the performance of the DASS model on weakly labeled audio-event classification tasks. We use the AudioSet dataset to train and evaluate our models. The dataset and the training details are described in the following sections.

\subsection{Dataset and Training Details}
AudioSet~\cite{gemmeke2017audio} contains over 2 million 10-second audio clips extracted from YouTube videos. These sounds clips are labeled from a set of 527 labels. The full training (AS-2M), balanced (AS-20K) and the evaluation dataset contain 2M, 20k and 22k data points respectively. We follow the training pipeline from AST~\cite{gong2021ast}, with a different learning rate. For both the balanced and full dataset we use a learning rate of 1e-4. For the balanced dataset, we train for 25 epochs and the learning rate is cut in half every 5 epochs after 10 epochs. For the full training set, we train the model for 10 epochs and cut the learning rate in half every epoch starting from the second epoch. We use the Adam optimizer with a batch size of 12 to train the model. 

We experiment with two different models: DASS-Small and DASS-Medium containing (2, 2, 8, 2) and (2, 2, 15, 2) layers in the four groups respectively. Both models have feature dimension of $C_1,C_2,C_3,C_4$ = (96,192,384,768) across the four groups. DASS-Small and DASS-Medium contain 30 and 49 million parameters. 

\subsection{Impact of Pretraining and Knowledge Distillation}

We compare ImageNet pretrained DASS with randomly initialized DASS with and without knowledge distillation. As observed in Table~\ref{tab:pretrainKD} the ImageNet pretrained models outperform the randomly initialized DASS models. 

Knowledge distillation improves the performance in both settings: when we use models pretrained on image data and when we do not. The performance improvement is higher when we do not have access to ImageNet pretrained models. 
We explored two different losses for knowledge distillation: KL-divergence and binary cross entropy. They performed similarly so we use KL-divergence for all experiments. 
To further explore the benefits of knowledge distillation, we continued training a randomly initialized DASS model for 100 epochs. This models achieves a mAP of 31.6 on the Audioset balanced set which is the same as AST with pretrained weights, and is close to the performance of DASS models with Imagenet pretraining and no knowledge distillation. DASS models trained with knowledge distillation outperform even the base AST model which has substantially more parameters than the DASS model.  

\begin{table}[h!]
    \centering
    \begin{tabular}{lccc} \toprule
         &  IN Pretrain &KD& mAP\\ \cmidrule{2-4}
         AST-Small (23M)& False & False & 10.6 \\
         AST-Base (86M) & False & False & 14.8 \\
         DASS-Small (30M)&  False &False& 12.0\\
         DASS-Small (30M)& False &True& \textbf{20.2} \\ \cmidrule{1-4}
         AST-Small (23M)& True & False & 31.0 \\ 
         AST-Base (86M)& True & False & 34.7 \\
         DASS-Small (30M)&  True & False & 34.6\\
         DASS-Small (30M)&  True &True& \textbf{38.4}\\
         \bottomrule
    \end{tabular}
    \caption{Comparison of DASS with AST on AS-20K. 
    Knowledge distillation (KD) is helpful with and without pretraining.}
    \label{tab:pretrainKD}
    \vspace{-10pt}
\end{table}

\subsection{Does a Stronger Teacher Makes Better Students}
To train the DASS model, we use the AST base model as the teacher. The DASS model outperforms the AST teacher model. In this section, we explore whether using a stronger DASS teacher would produce a stronger student.  To reduce computational costs, we limit  training to 5 epochs. As shown in Table~\ref{tab:DASS_teacher}, using a stronger teacher does not result in a better performing student, although it is possible the student benefits more from a teacher with a different architecture. Previous studies have observed similar trends for CNN/Transformer based teacher-student models~\cite{gong2022cmkd}.

\begin{table}[h!]
    \centering
    \begin{tabular}{llcc} \toprule 
        \multicolumn{2}{c}{model} &  \multicolumn{2}{c}{mAP} \\ \midrule
         Student & Teacher & Teacher & Student \\ \midrule 
        DASS-Small & AST-Base & 45.9 & 47.1 \\
        DASS-Medium & AST-Base & 45.9 & 47.1 \\
        DASS-Small & DASS-Small & 47.1 & 46.9 \\
        DASS-Small & DASS-Medium & 47.1 & 46.7 \\
        DASS-Medium & DASS-Small & 47.1 & 47.1 \\
        DASS-Medium & DASS-Medium & 47.1 & 46.9 \\ \bottomrule
    \end{tabular}
    \caption{Performance comparison on AS-2M for different teachers for DASS student models}
    \label{tab:DASS_teacher}
    \vspace{-10pt}
\end{table}

\begin{table}[h!]
    \centering
    \begin{tabular}{lccc} \toprule
        & Params & Pretrain & mAP \\ \midrule 
        \multicolumn{3}{l}{\textbf{Transformer based models}} \\ 
         AST~\cite{gong2021ast} & 87M & IN SL & 45.9\\
        HTS-AT~\cite{chen2022hts} & 31M & IN SL & 47.1\\ 
        PaSST~\cite{koutini2021efficient} & & IN SL & 47.1\\ 
        Audio-MAE$\dagger$~\cite{huang2022masked} & 86M & SSL & 47.3 \\
        BEATs(iter3)~\cite{chen2022beats} & 90M & SSL & 48.6 \\
        EAT~\cite{chen2024eat} & 88M & SSL & 48.6 \\ \midrule 
        \multicolumn{3}{l}{\textbf{Concurrent SSM models}} \\ 
        AuM~\cite{hamza2024audio} & 26M & IN SL & 39.7 \\ 
        Audio Mamba~\cite{lin2024audio} & 40M & IN SL  & 44.0\\ 
        DASS-Small & 30M & IN SL & 47.2\\ 
        DASS-Medium & 49M & IN SL & {47.6}\\ \midrule 
        \multicolumn{3}{l}{\textbf{Teacher ensemble: AST + HTS-AT}} \\ 
        DASS-Small & 30M & IN SL & 48.6\\ 
        DASS-Medium & 49M & IN SL & \textbf{48.9}\\ \bottomrule
        
    \end{tabular}
\caption{Performance comparison on AS-2M. IN SL: ImageNet supervised learning; SSL: self-supervised learning; $\dagger$ Industry-level computation}
    \label{tab:main_mAP}
    \vspace{-10pt}
\end{table}

\begin{figure*}[th!]
    \centering
    \includegraphics[width=2\columnwidth]{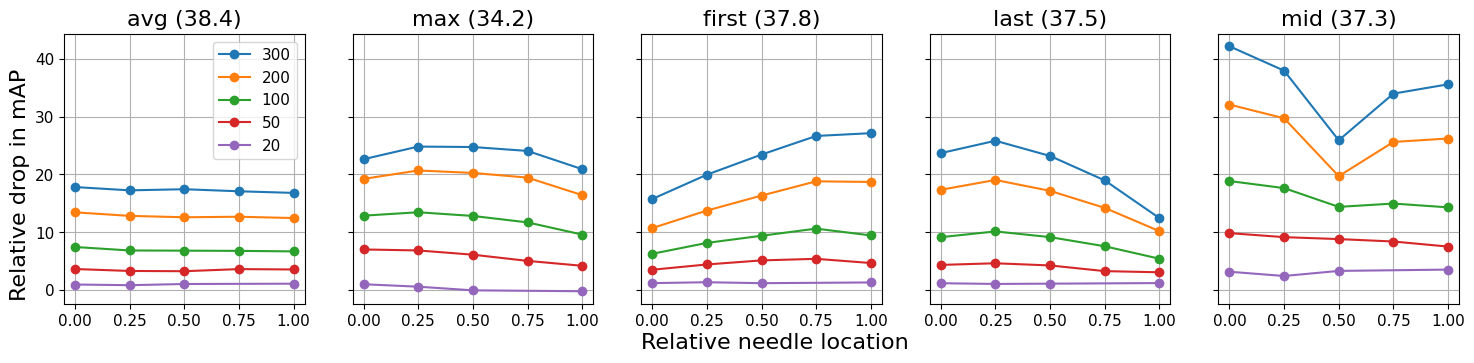}
    \caption{Pooling strategy performance on the NIAH task for DASS-Small trained on AS-20K. Parenthesis shows 10s performance.}
    \label{fig:NIAH_vs_pool}
    \vspace{-10pt}
\end{figure*}

\subsection{AudioSet: Comparison with Other Methods}

We compare our proposed DASS models with both transformer-based state-of-the-art and concurrent SSM-based methods.  As observed in Table~\ref{tab:main_mAP}, DASS outperforms all the existing methods. 
DASS has fewer parameters than AST and audio-MAE (86M vs 49M) and requires significantly lower computational resources to train (64 V100s for audio-MAE vs 1 A5000 for DASS). DASS and other SSM-based models also have faster inference and require less computational resources during inference than transformer-based models. 

By using knowledge distillation to train the SSMs, we bridge the performance gap between the transformer and SSMs.  DASS provides the best of both worlds: it outperforms the state-of-the-art transformer-based models yet has the efficiency and ability to process long audio sequences.

\section{Audio Needle in a Haystack: \\evaluation on longer sequences}
One of the main advantages of SSMs is their ability to handle longer sequences~\cite{gu2023mamba} with linear complexity, unlike Transformers. Recent state-space-based models show remarkable performance on the audio-event classification task. Some of them simulate longer input sequences to compare speed and GPU memory usage with transformer-based models such as AST~\cite{huang2022masked,hamza2024audio}. Although their brief analysis shows that state-space models can theoretically handle much longer input sequences than AST, they do not measure the performance of state-space models at these lengths. 

One of the objectives for our SSM research was to experimentally quantify the duration scalability of the DASS models. To do this we train the DASS only on the 10-second spectrograms. During evaluation, we synthesize longer input sequences to measure the duration robustness of the model.

We design an Audio Needle in A Haystack (Audio NIAH) task where we randomly insert a needle: a 10-second audio spectrogram from the Audioset dataset into a larger haystack of various lengths. We construct the haystack in two ways: first by zero padding to achieve the desired length and second, by appending noise at different levels. For the latter case, we generate noise waveforms at various SNRs and then generate filterbank (fbank) features using the same pipeline used for generating fbank features for the needle. In both cases, the needle (10-second audio spectrogram) is not modified.  As mentioned previously, the models are trained only on the 10-second spectrogram, and the NIAH test is performed during evaluation to measure the duration robustness of the model.

\subsection{NIAH: Impact of Pooling Method }
Our first NIAH experiment explored whether the location of the needle results in a performance difference, or whether the DASS model can pick up information from anywhere in the haystack. To do so, we create a haystack of various lengths and insert the 10s spectrogram at different relative positions. We insert the needle at 0 ( the beginning), 0.25, 0.5 ( the middle), 0.75 and 1.0 (the end) in the haystack.  We measure the impact of needle position by measuring the relative drop w.r.t to the base performance of the DASS model i.e. when we do not use a haystack and just use the 10s spectrogram as input to the DASS model as: $(mAP_{10} - mAP_t) / (mAP_{10})$.

The AST model relies on a CLS token to summarize information, whereas SSMs offer a natural way to summarize information into a single embedding. For example, the last embedding summarizes all the information seen so far. The bidirectional nature allows us to use embeddings at any time step as the CLS embedding.  

DASS uses a pooling method to generate the CLS token and summarize information before the classifier. We explore the impact of the pooling strategy on the NIAH task. As seen in Figure~\ref{fig:NIAH_vs_pool}, the choice of pooling method affects the performance depending on the relative needle location. 

When the model is trained with the first token embedding as the CLS token, it performs better on the NIAH task when the needle is in the beginning. Similarly, DASS models trained with last and mid as CLS tokens have the lowest drop in performance when the needle position is last or in the middle of the haystack respectively. Average pooling shows the least sensitivity to the relative position of the needle position. We also explored sum as the pooling mechanism but it resulted in severe drops in performance especially at longer haystack durations. 

For all pooling methods, the performance drop increases with an increase in the haystack length.  This result makes sense as the model has to process and summarize longer and longer utterances. Although, for the DASS model the performance drop at 50 second haystack length, is less than 5\% whereas AST loses most of its performance at these lengths.

\begin{figure}[h!]
    \centering
    \includegraphics[width=0.9\columnwidth]{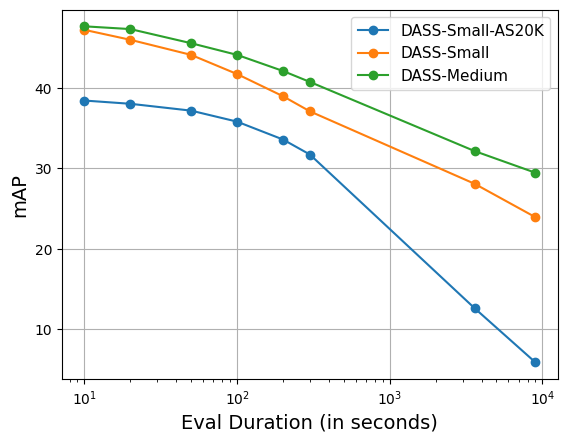}
    \caption{mAP performance across different hay stack lengths. Haystack: zero padding, needle location: 0.5}
    \label{fig:mAP_vs_dur_0pad}
    \vspace{-12pt}
\end{figure}

\subsection{NIAH: Impact of Haystack Length on Performance}
Our second NIAH experiment explored the limits of the DASS model. For this experiment, we fixed the needle position to the middle of the haystack and increased the haystack length from 10 seconds to up to 2.5 hours (a 900-fold increase in the evaluation duration compared to the training duration). We conduct the 2.5-hour experiment on a single A6000 GPU. 

As observed in Figure~\ref{fig:mAP_vs_dur_0pad}, the stronger models show more robustness to changes in length. The small model trained on the smaller balanced subset drops significantly in performance compared to the small model that was trained on the full AudioSet. The medium model trained on the full AudioSet, which performs similarly to the small model, is even more robust to changes in evaluation length. The over-parametrization helps the DASS models be more duration scalable. The medium model at extreme length (i.e. 2.5 hours) performs similarly to a small model trained on the balanced subset. 

We carry out a similar experiment on the transformer-based AST model. To avoid the mismatch between the learned positional embeddings during training and testing, we train an AST model with sinusoidal embedding. This allows us to extend the inputs and positional embeddings to arbitrary lengths.  For AST, we place the needle only at the beginning of the haystack so there is no mismatch between the positional embeddings added to the needle during training and testing. 

The AST model loses performance rapidly with an increase in the haystack length. At only 30 seconds evaluation length, the performance drops to 20\% of the performance at 10 seconds length which does not happen to DASS models even at 2.5 hours evaluation length. We believe this is due to attention weights becoming very small by being spread over more tokens. In the future, we would like to analyze the attention maps of AST under various evaluation durations. 

These results are particularly useful since we do not need to train the DASS model on longer input sequences thereby reducing the overall computational resources required to train and use these models for longer data modalities such as videos. 
\begin{figure}[t]
    \centering
    \includegraphics[width=0.9\columnwidth]{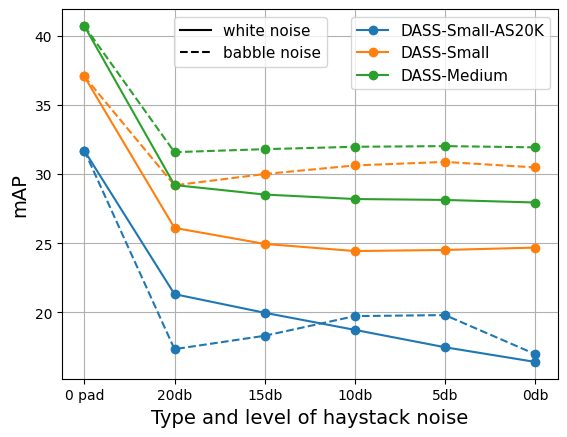}
    \caption{mAP vs type and level of white or babble noise in the haystack. Haystack length: 300 seconds}
    \label{fig:NIAH_mAP_vs_snr}
    \vspace{-10pt}
\end{figure}

\begin{figure}[b!]
    \centering
    \includegraphics[width=0.9\columnwidth]{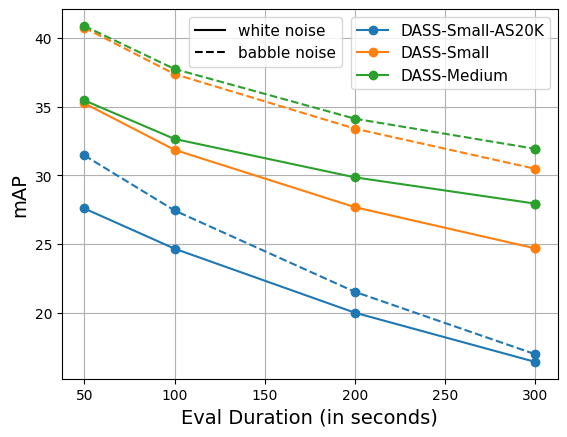}
    \caption{mAP performance across different haystack lengths. Haystack: white noise at 0 SNR, needle location: 0.5.}
    \label{fig:NIAH_mAP_vs_dur_0snr}
    \vspace{-12pt}
\end{figure}

\subsection{NIAH: Impact of Haystack Construction}
As mentioned earlier, we construct the haystack in two ways: by zero padding and by padding noise. Here we explore how DASS models perform under two different haystack constructions. We explore two types of noise: Gaussian noise and babble noise. The babble noise is constructed by combining 30 audio babble noise files from the MUAVIC dataset~\cite{anwar2023muavic}. 

For both noises, the DASS models perform worse than zero padded haystacks. For babble noise, we see mixed results, the performance goes up when the SNR goes down and then again decreases at 0 SNR. For white noise, the performance goes down when going from 20 SNR to 15 but stays relatively the same after that. 

We believe this result is because the AudioSet dataset contains speech and white noise as one of the classes and the DASS model might be classifying the haystack noise correctly as speech or noise which is affecting the performance. We observe more variation in performance for babble noise since AudioSet contains significantly more data points labeled as speech than white noise. For both noise settings, the stronger models show more robustness to noise as well. 

To measure the impact of duration on the noise-based haystack, we fix the needle location to the middle, fix the noise SNR to 0, and vary the length of the haystack. As seen in Figure~\ref{fig:NIAH_mAP_vs_dur_0snr}, the performance goes down with an increase in length. The drop in performance is significantly more for noise-based haystacks than for zero-padded haystacks. The DASS models at 300 seconds with a noise-based haystack perform worse than 2.5 hours of haystack with zero padding.

\section{Conclusions}
In this paper, we propose DASS: a state space-based model that achieves state-of-the-are results for audio event classification. DASS has the performance of compute-heavy transformer-based models and the efficiency of the state-space methods. Our experiments show that knowledge distillation is helpful across dataset size, and model scale. DASS outperforms the AST-based teacher. 

Audio SSM can support much longer utterances theoretically.  We propose an Audio NIAH task to measure the impact of duration mismatches between the training and evaluation lengths on the performance.  DASS is significantly more robust to duration mismatches during training and evaluation compared to AST. We also observe that stronger DASS models tend to be more robust to duration mismatches. 

\bibliographystyle{IEEEbib}
\bibliography{refs}

\end{document}